\documentclass[aps,prl,preprint,groupedaddress]{revtex4-1}
\usepackage{graphicx}
\usepackage{color}
\usepackage{epstopdf, epsfig}
\usepackage{subfig}

\begin{document}


\title{Length scales and eddy viscosities in turbulent free shear flows}


\author{Gioacchino Cafiero}
\email[]{gioacchino.cafiero@polito.it, Dipartimento di Ingegneria Meccanica e Aerospaziale, Politecnico di Torino, Italy}
\affiliation{Turbulence, Mixing and Flow control group, Dept of Aeronautics, Imperial College London (UK)}
\author{Martin Obligado}
\email[]{martin.obligado@univ-grenoble-alpes.fr}
\affiliation{Univ. Grenoble Alpes, CNRS, Grenoble INP, LEGI, 38000 Grenoble, France}
\author{John Christos Vassilicos}
\email[]{j.c.vassilicos@imperial.ac.uk}
\affiliation{Turbulence, Mixing and Flow control group, Department of Aeronautics, Imperial College London (UK)}

\date{\today}

\begin{abstract}
In the present paper, we address the important point of the
proportionality between the longitudinal integral lengthscale ($L$)
and the characteristic mean flow width ($\delta$) using experimental
data of an axisymmetric wake and a turbulent planar jet. This is a
fundamental hypothesis when deriving the self-similar scaling laws in
free shear flows, irrespective of turbulence dissipation scaling.  We
show that $L/\delta$ is indeed constant, at least in a range of
streamwise distances between 15 and 50 times the characteristic inlet
dimension ($L_{ref}$, nozzle width or wake generator
size). Furthermore, we revisit turbulence closure models such as the
Prandtl mixing length \cite{prandtl1925} and the constant eddy
viscosity in the light of the non-equilibrium dissipation scalings. We
show that the mixing length model, with $l_m\sim \delta$, does not
comply with the scalings stemming from the non-equilibrium version of
the theory; we instead obtain $l_m\sim \delta
\sqrt{Re_G/Re_{0\delta}}$, where $Re_G$ and $Re_{0\delta}$ are a global
and local Reynolds number, respectively. Similarly, the eddy viscosity
model holds in the case of the non-equilibrium version of the theory
provided that the eddy viscosity is constant everywhere, not only
across sections orthogonal to the streamwise direction as in the
equilibrium case. We conclude comparing the results of the different
models with each other and with experimental data and with an improved
model (following Townsend) that accounts for the intermittency of the
flow and corrects for the eddy viscosity variation across the flow
boundaries.

\end{abstract}


\maketitle

\section{Introduction}

Free shear flows are of significant importance in many natural and
industrial applications. They are also of great interest for
fundamental research, as it is one of the few cases in turbulence
where mean quantities can be predicted under a small, physically
based, set of hypotheses. The theory, derived by Townsend
\citep{Townsend} and later by George \citep{george1989}, requires the
self-preservation of some turbulence quantities in the mean momentum
and streamwise kinetic energy equations. In order to close the
equations, an {\it{ad hoc}} assumption is required to model the
dissipation term in the kinetic energy equation. The closure usually
chosen is the one consistent with the Richardson-Kolmogorov
cascade. One assumes that the centreline turbulence dissipation rate
$\varepsilon$ can be described as $\varepsilon=C_\varepsilon
K^{3/2}/L$, where $K$ is the turbulent kinetic energy, $L$ is an
integral length-scale of turbulence (usually taken to be the
longitudinal one) and $C_\varepsilon$ is a dimensionless coefficient
which may depend on boundary conditions but it is independent of
Reynolds number at high enough Reynolds number values. Finally, the
integral lengthscale $L$ is assumed to be proportional to a mean flow
profile width such as the wake/jet width $\delta$.

Focusing on the free shear flows that will be investigated in the
present work, namely the axisymmetric wake and the planar jet, this
theoretical approach leads to the following streamwise evolutions of
the centreline velocity (jet) or velocity deficit (wake) $u_0$ and the
jet or wake width $\delta$: 
\begin{equation}
u_0 \sim  (x-x_0)^a,
\label{eq:uo}
\end{equation}
\begin{equation}
\delta \sim  (x-x_0)^b,
\label{eq:delta}
\end{equation}
with $a=-1/2$ and $b=1$ for the planar jet and $a=-2/3$ and $b=1/3$ for the axisymmetric wake \cite{Townsend}, \cite{george1989}, \cite{pope} ($x$ is the streamwise coordinate and $x_0$ is a virtual origin). 

Previous to Townsend \cite{Townsend} and George \cite{george1989},  researchers were already able to predict these exact same streamwise dependencies of $\delta$ and $u_0$ under different assumptions. A closure of the mean momentum equation can be given by assuming that the relevant component of the Reynolds shear stress tensor, $<u^{\prime}v^{\prime}>$, is related to the streamwise mean flow velocity $\overline{u}$ by 
\begin{equation}
{<u'v'>}=-\nu_T \frac{\partial{\overline{u}}}{\partial{y}}
\label{eq:eddyvisc}
\end{equation}
\noindent where $y$ is the spreading direction of the flow and $\nu_T$ is the eddy viscosity. The modelling of $\nu_T$ has been the focus of intense research during the first half of the 20th century (\cite{chou}, \cite{goldstein}, \cite{townsend1949}, \cite{corrsin}) and has been studied for many free shear flows. The most basic and common hypotheses used are Prandtl's mixing length hypothesis,  
\begin{equation}
\nu_T=l_m^2\frac{\partial{\overline{u}}}{\partial{y}},
\label{eq:Mixinglength}
\end{equation}
\noindent where $l_m$ is the mixing length, and a constant eddy viscosity hypothesis
\begin{equation}
\nu_T= L_{eddy}U_{eddy}
\label{eq:EddyViscosity}
\end{equation}
\noindent where $L_{eddy}$ and $U_{eddy}$ are characteristic scales of length and velocity, respectively, which may depend on $x$ but are constant along $y$.

Equations \ref{eq:uo} and \ref{eq:delta} are retrieved with $l_m \sim
\delta$ and $L_{eddy} \sim \delta$. In this sense, the
Richardson-Kolmogorov cascade (which is one of the pillars of the
Townsend-George approach given that it adopts $\varepsilon \sim
K^{3/2}/\delta$) is consistent with both the constant eddy viscosity
and the Prandtl's mixing length hypotheses.

Recent works have unveiled the presence of turbulence dissipation
scalings in free shear flows which are at odds with the
Richardson-Kolmogorov scaling for $\varepsilon$ (\cite{cvjet},
\cite{dairay2015}, \cite{obligado2016}, \cite{ZV2017}). Direct
numerical simulations (DNS) and experiments for axisymmetric wakes and
experimental data for the plane jet suggest the presence of free shear
flows that follow non-equilibrium scalings for dissipation, at least
for a large portion of the flow. The dissipation parameter
$C_\varepsilon$ is no longer independent of Reynolds number (though it
does remain independent of the fluid's kinematic viscosity $\nu$), but
scales as $Re_G/Re_{\delta}$ where $Re_G$ $= U_{ref} L_{ref}/\nu$, ($
U_{ref}$ being the free stream or inlet velocity and $L_{ref}$ a
characteristic inlet lengthscale such as the wake generator's size or
the nozzle width) is the global Reynolds number and $Re_{\delta} =
\sqrt{K_0} \delta/\nu$ ($K_0$ being the turbulent kinetic energy on
the flow centreline) is a local Reynolds number. As evidenced by
Dairay et al \cite{dairay2015} and Cafiero and Vassilicos
\cite{cvjet}, the application of the non-equilibrium dissipation
scaling, makes it possible to use a smaller number of assumptions than
Townsend \cite{Townsend} and George \cite{george1989} and leads to new
exponents for eq. (\ref{eq:uo}) and (\ref{eq:delta}). In the planar jet
case, $a=-1/3$ and $b=2/3$ while in the axisymmetric wake case, $a=-1$
and $b=1/2$. It is important to explicitly notice that both in the
classical and the non equilibrium dissipation scaling the assumption
$L\sim \delta$ is needed. In table \ref{tab:theory} we summarise the
scalings stemming from the Richardson-Kolmogorov equilibrium
dissipation and the non-equilibrium dissipation versions of the
theory.

\begin{table}
  \begin{center}
\def~{\hphantom{0}}
  \begin{tabular}{lccccc}
                    		 					&&& Equilibrium	                 & & Non-equilibrium\\[3pt]
      Dissipation Scaling    					&&&  $K_0^{3/2}/\delta$  & &      $(Re_G/Re_\delta )^{m} K_{0}^{3/2}/\delta$	   	  \\
      Power laws exponents: Axisymmetric Wake   &&& $a=-2/3$, $b=1/3$    & &      $a=-1$, $b=1/2$  \\
      Power laws exponents: Planar Jet          &&& $a=-1/2$, $b=1$      & &        $a=-1/3$, $b=2/3$  \\	     
      Mixing length $l_m$          				&&& $\sim \delta$        & &        $\sim \delta \sqrt{Re_G/Re_{0\delta}}$  \\	
      Eddy viscosity $\nu_T$          			&&& $\sim u_0 \delta$    & &        $\sim U_{ref}L_{ref}$  \\     
  \end{tabular}
  \caption{Summary of the mean flow ($u_0$), characteristic flow width
    ($\delta$), mixing length ($l_m$) and eddy viscosity ($\nu_T$)
    scalings obtained according to the equilibrium and the
    non-equilibrium versions of the Townsend-George theory for the
    axisymmetric wake and turbulent planar jet cases.}
  \label{tab:theory}
  \end{center}
\end{table}

One easily checks that the Prandtl mixing length hypothesis cannot
lead to (\ref{eq:uo}) and (\ref{eq:delta}) with non-equilibrium
exponents if $l_m \sim \delta$. The non-equilibrium scalings can
however be retrieved if $l_m \sim \delta \sqrt{Re_G/Re_{0\delta}}$
where $Re_{0\delta} = u_{0}\delta/\nu$ (see table \ref{tab:theory}).
As for the constant eddy viscosity, it can still be used to obtain
non-equilibrium scaling exponents $a$ and $b$ but only if $\nu_T$ is
constant throughout the flow so that $\nu_T\sim U_{ref} L_{ref}$, not
only across sections of the flow orthogonal to the streamwise
coordinate as in $\nu_T\sim u_0 \delta$.

In this work we first and foremost address one key aspect of the
Townsend-George approach using experimental data for a turbulent
axisymmetric wake at $Re_G=40000$ and a planar jet at $Re_G=20000$:
the important question of the proportionality of $L$ and $\delta$ in
free shear flows. Secondly, we also ask whether the equilibrium and
non-equilibrium mixing length and eddy viscosity models imply
different mean flow profiles and how the different models perform in
predicting these profiles.


\section{Experimental setup}
\begin{figure}
\centering
{\includegraphics[width=\columnwidth]{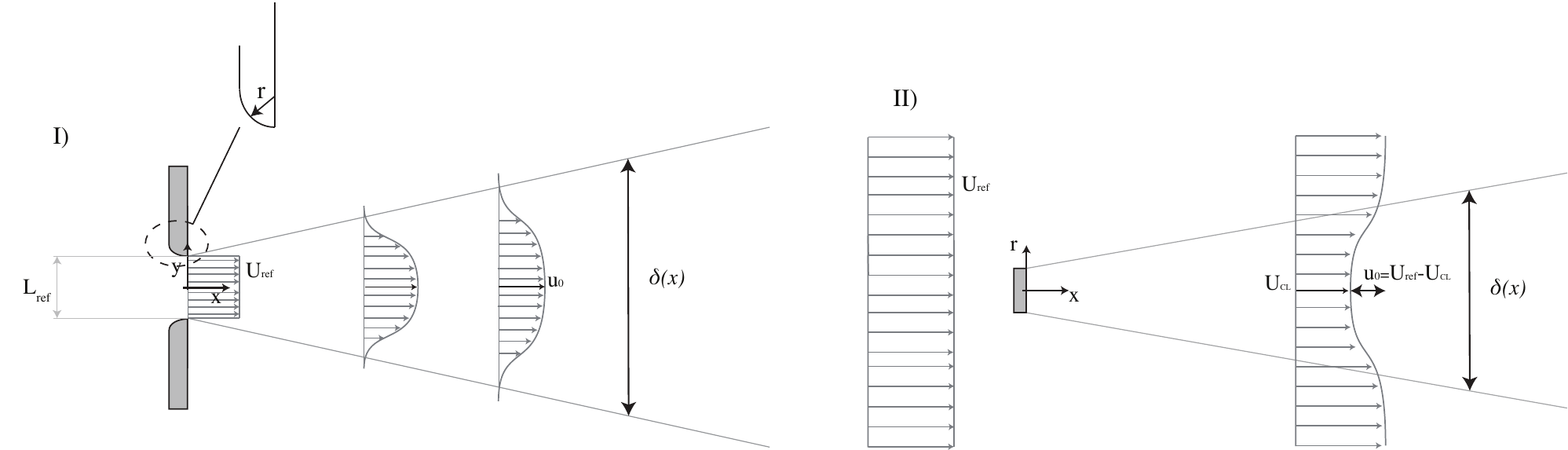}}
\caption{I) Schematic representation of the planar jet flow; II) Schematic representation of the axisymmetric wake flow.  $\delta(x)$ is representative of the characteristic flow width at each streamwise location $x$. }
\label{fig:expsetup}
\end{figure}
The experiments are carried out in two different facilities at Imperial College London. A schematic representation of both the planar jet  and the axisymmetric wake flows is provided in figure \ref{fig:expsetup}(I) and \ref{fig:expsetup}(II), respectively. 
The planar jet flow is generated using a centrifugal blower which collects air from the environment and then forces it into a plenum chamber. In order to reduce the inflow turbulence intensity level and remove any bias due to the feeding circuit, the air passes through two sets of flow straighteners before entering a convergent duct (having area ratio equal to about 8). At the end of the duct there is a letterbox slit with aspect ratio $w/L_{ref}=31$ (with $L_{ref}=15$ mm).  In order to produce a top hat entrance velocity profile, the two longest sides of the slit are filleted with a radius $r=2L_{ref}$, following the careful recommendation by \cite{deo}. The jet exhausts in still ambient air and is confined in the spanwise direction by two perspex walls of size $100L_{ref} \times 100L_{ref}$ placed in $x-y$ planes ($L_{ref}$ is the slot width in the $y$ direction). The rotational speed of the blower is controlled using an in-house PID controller to produce an inlet Reynolds number $Re_G=20000$. Single (SW) and Cross (XW) wire measurements are taken along the jet centreline in the range $x/L_{ref}=14-50$ with 2$L_{ref}$ spacing. Both SW and XW are driven by a Dantec Streamline constant temperature anemometer (CTA). Data are sampled at a frequency of 50 KHz, with measurements lasting 60 s and 120 s respectively in the SW and XW cases.  

The wake flows are generated in the low-turbulence wind tunnel at Imperial College London. The measurement test section is 3 ft x 3 ft ($\approx $91 cm x 91 cm) and length 4.25 m. The plates employed for these experiments have a reference length  $L_{ref}=\sqrt{A_{plate}}=$  64 mm, with thickness 1.25 mm , ${A_{plate}}$ being the frontal area of the plate. The plate is suspended in the centre of the wind tunnel normal to the laminar free stream using four 1 $mm$ diameter piano wires. The free-stream velocity was kept fixed at $U_{ref}$=10 m/s using a PID controller. For that value, the velocity fluctuations around the mean are below 0.1\% when the plate is not in place. The velocity signal is measured using a one component hot-wire (herein referred to as SW) driven by a Dantec Streamline constant temperature anemometer (CTA). Data are sampled at a frequency of 20 KHz . Each measurement lasts for 60 s, which was deemed to be sufficient to converge the integral scales. {Finally, a X-wire probe was used to estimate the kinetic energy only for centreline measurements. The centreline kinetic
energy is calculated  by assuming axial symmetry, {i.e.}  $K_0=0.5 \left(<u_{x}^{\prime^{2}}>+2<u_{r}^{\prime^{2}}>\right)$ where $u_{x}^{\prime}$ and $u_{r}^{\prime}$ are streamwise and radial fluctuating velocities. More details about the experimental set-up can be found in \cite{dairay2015, obligado2016}.

The longitudinal integral lengthscale $L$ is calculated converting the anemometer time signal into space using the frozen turbulence hypothesis and using the autocorrelation of the fluctuating streamwise velocity. Comparison of the results with those obtained using the expression proposed by \cite{tennekes&lumley}, \textit{i.e.} $L=\pi E_{u}(k=0)/u'^2$, shows minimum discrepancies. The estimate of the turbulent dissipation rate $\varepsilon$ is obtained from its isotropic surrogate, i.e. $\varepsilon_{ISO}=15 \nu \overline{(\partial u^{\prime}/\partial x)^2}$, by integrating the one dimensional spectrum of the velocity signal $F_{11}^{(1)}$ as follows
\begin{equation}
 \overline{(\partial u^{\prime} /\partial x)^2} = \int_0^{\infty} k^2 F_{11}^{(1)} dk . 
\label{eq:2ndDer}
\end{equation}
In both the experiments, we took care of reducing the noise at the high wavenumber end of the spectrum.  As suggested by \cite{obligado2016}, we fit the portion of the spectrum at frequencies higher than Kolmogorov's frequency with an exponential law. It must be however remarked that the contribution of this portion of the spectrum to the integral in equation (\ref{eq:2ndDer}) is always less than 6\%.

\section{Results} \label{sec:results}
\begin{figure}
\centering
\subfloat[][]
{\includegraphics[width=0.5\columnwidth]{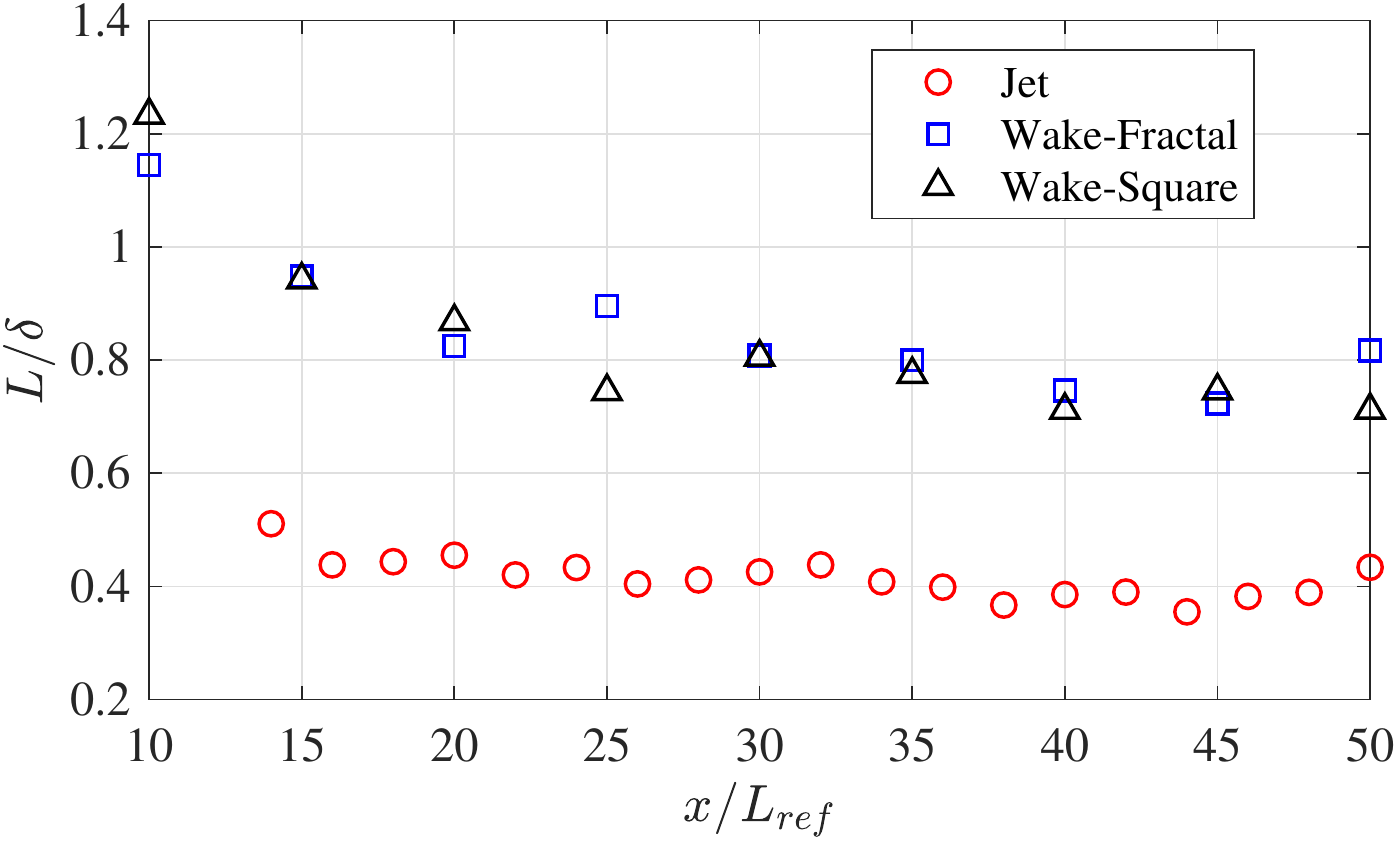}}
\subfloat[][]
{\includegraphics[width=0.5\columnwidth]{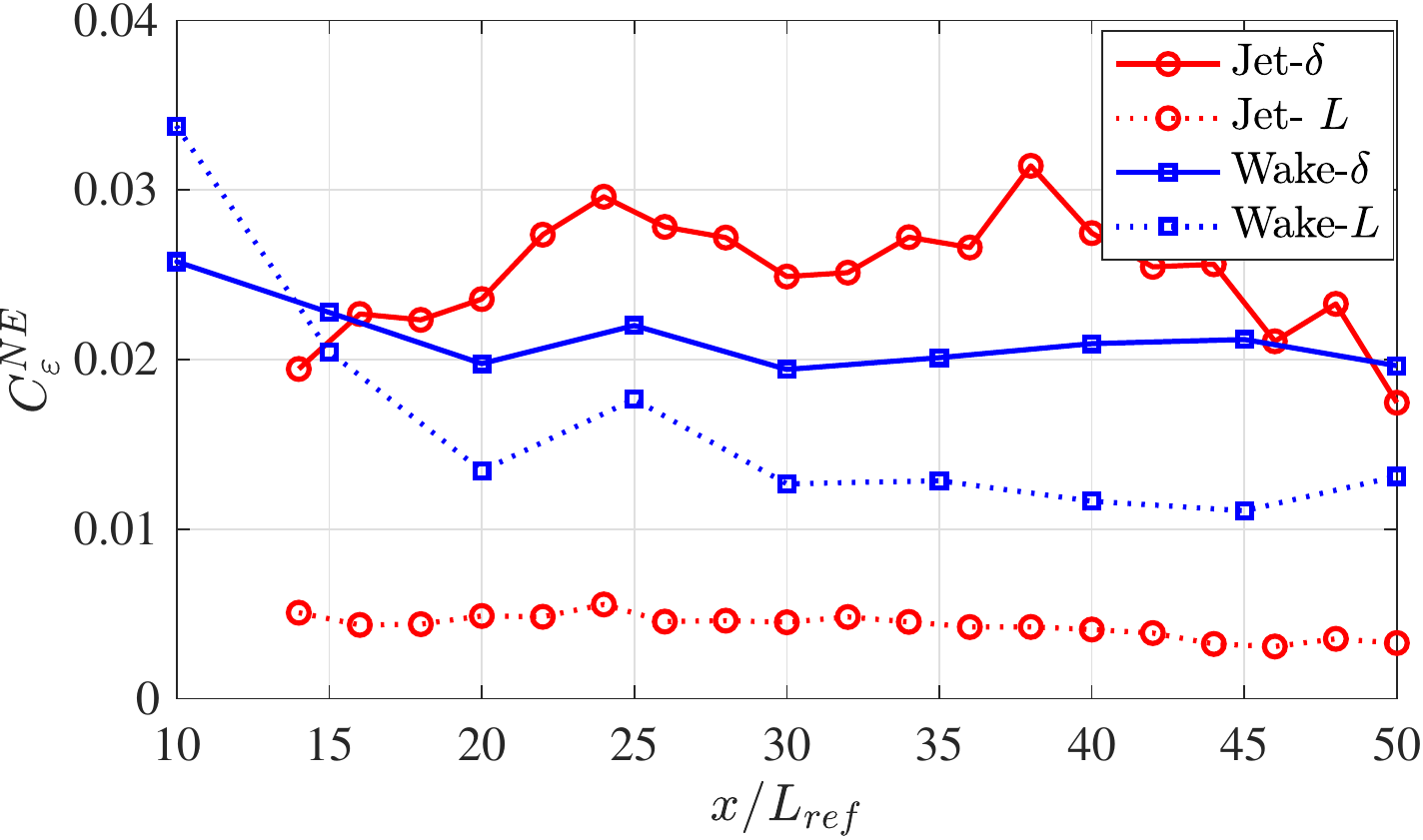}}\\
\caption{a){Streamwise profiles of $L/\delta$} for the planar jet (red circles), the axisymmetric fractal wake (blue squares) and the axisymmetric square wake (black triangles). The ratio is calculated along the centreline of the flow. Inlet Reynolds numbers $Re_G=40000$ (wakes) and $20000$ (jet).  b) Non-equilibrium dissipation constant $C_{\varepsilon}^{NE,\delta} =  (U_{ref} L_{ref})  \varepsilon  \delta^2/K_0$ for the the planar jet (red circles-continuous) and the axisymmetric fractal wake (blue squares-continuous). Non-equilibrium dissipation constant $C_{\varepsilon}^{NE,L} =(U_{ref} L_{ref})  \varepsilon L^2/K_0$ for the planar jet (red circles-dotted) and the axisymmetric fractal wake (blue squares-dotted). }
\label{fig:ldeltadiff}
\end{figure}

Townsend \cite{Townsend}, George \cite{george1989}, Dairay et al
\cite{dairay2015} and Cafiero \& Vassilicos \cite{cvjet} assumed that
the dissipation lengthscale $C_\varepsilon K^{3/2}/\varepsilon$ could
be interchangeably taken to be proportional to the integral
lengthscale $L$ or the characteristic flow width $\delta$ without loss
of validity of their results, at least in terms of scaling. It is then
pertinent to investigate this assumption by looking at data measured
for two different wake generating bodies, a square and a fractal (see
\cite{dairay2015,obligado2016} for more details), as well as for the
planar jet. Data are taken along the flows' centreline. The inlet
Reynolds numbers are $Re_G=40000$ and $20000$ for the wakes and jet,
respectively.  Figure \ref{fig:ldeltadiff}a supports the assumption of
proportionality between $L$ and $\delta$, at least in the range of
streamwise distances $15 \leq x/L_{ref} \leq 50$, which contains the
region where the non-equilibrium dissipation scaling holds as reported
in \cite{dairay2015, obligado2016, cvjet}. A constant value of the
ratio $L/\delta$ is attained both in the wake and the jet cases, but
the value of the constant seems to vary from flow to flow. In general,
there is no reason to expect any sort of universality for this
ratio. For example, the planar jet is characterized by larger
entrainment, thus entailing higher spreading rate and flow width
$\delta$.

Figure \ref{fig:ldeltadiff}b compares the profiles of
$C_{\varepsilon}^{NE}$ (defined in the caption of figure
\ref{fig:ldeltadiff}) as a function of the streamwise distance when
calculated using either $L$ or $\delta$. We plot the data for the
turbulent planar jet and the axisymmetric wake of the fractal obstacle
only, as the square one exhibits similar trends. Regardless the choice
of $L$ or $\delta$, $C_{\varepsilon}^{NE}$ still exhibits the same
constant behaviour, as required by the non-equilibrium turbulence
dissipation scaling.

\subsection{Lateral Profiles}\label{sec:prof}
In this section we investigate the consequences of the application of
a different turbulence dissipation scaling on the lateral mean flow
profiles. The scalings (\ref{eq:uo}) and (\ref{eq:delta}) and stemming
from the non-equilibrium version of the theory, can also be obtained
with the mixing length,
\begin{equation}
l_m \sim \delta \sqrt{Re_G/Re_{0\delta}},
\end{equation} 
but not with $l_m \sim \delta$ (see table \ref{tab:theory}). 

Similarly, for the constant eddy viscosity model, the non-equilibrium version of the theory returns the right exponent $a$ and $b$ provided that $\nu_T$ is not only constant across a section orthogonal to the mean flow as in the equilibrium case, but throughout. Introducing into equation (\ref{eq:eddyvisc}) the scalings of $\langle u^{\prime}v^{\prime}\rangle$ stemming from the non-equilibrium version of the theory theory (see \cite{dairay2015} and \cite{cvjet}), we obtain
\begin{equation}
- \nu_T \frac{\partial \overline{u}}{\partial y} = u_0^2 \frac{d\delta}{dx},
\end{equation} 
for the planar jet case \cite{cvjet} and
\begin{equation}
- \nu_T \frac{\partial \overline{u}}{\partial y} = U_{ref} u_0\frac{d\delta}{dx},
\end{equation} 
for the axisymmetric wake case \cite{dairay2015}, \cite{obligado2016}. Introducing the power laws (\ref{eq:uo}) and (\ref{eq:delta}) with the non-equilibrium values of $a$
and $b$ reported in table \ref{tab:theory},

\begin{equation}
\nu_T \sim U_{ref}L_{ref} \sim const,
\end{equation} 
both for the planar jet and for the axisymmetric wake case, as opposed to 
\begin{equation}
\nu_T \sim u_0\delta,
\end{equation} 
\noindent obtained from the equilibrium version of the Townsend-George
theory (which actually requires one more assumption to conclude, see
\cite{dairay2015} and \cite{cvjet}).

It is then important to determine whether the differences in mixing
length and in the eddy viscosity reflect in different mean flow
profiles for different turbulent dissipation scalings. Furthermore, it
is also important to determine whether the mean flow profiles obtained
with the Prandtl mixing length or with the constant eddy viscosity
models are consistent with the experimental data, and more or less so
depending on turbulence dissipation scaling.

Mixing-length based models \cite{corrsin, goldstein} have largely
proven to be inadequate for correctly predicting lateral mean flow
profiles. Townsend \cite{townsend1949}, comparing with the
experimental results obtained in the turbulent planar wake of a square
cylinder, finds that a constant eddy viscosity $\nu_T \sim u_0 \delta$
best represents his results. We also aim at comparing the mixing
length based model with a constant value of the eddy viscosity, but by
taking into account the non-equilibrium modification of these two
models.  Furthermore, following Townsend's approach
\cite{townsend1949}, we also correct the eddy viscosity to account for
the intermittency of the flow. In the following we compare our
experimental data with mean flow profiles predicted by Prandtl's
mixing length and constant eddy viscosity models for the axisymmetric
wake and the planar jet.

The detailed derivations of the profiles stemming from these two models modified to take into account the non-equilibrium cascade are reported in the appendix. 
\subsubsection{Axisymmetric Wake}
\begin{figure}
\centering
\subfloat[][]
{\includegraphics[width=0.5\columnwidth]{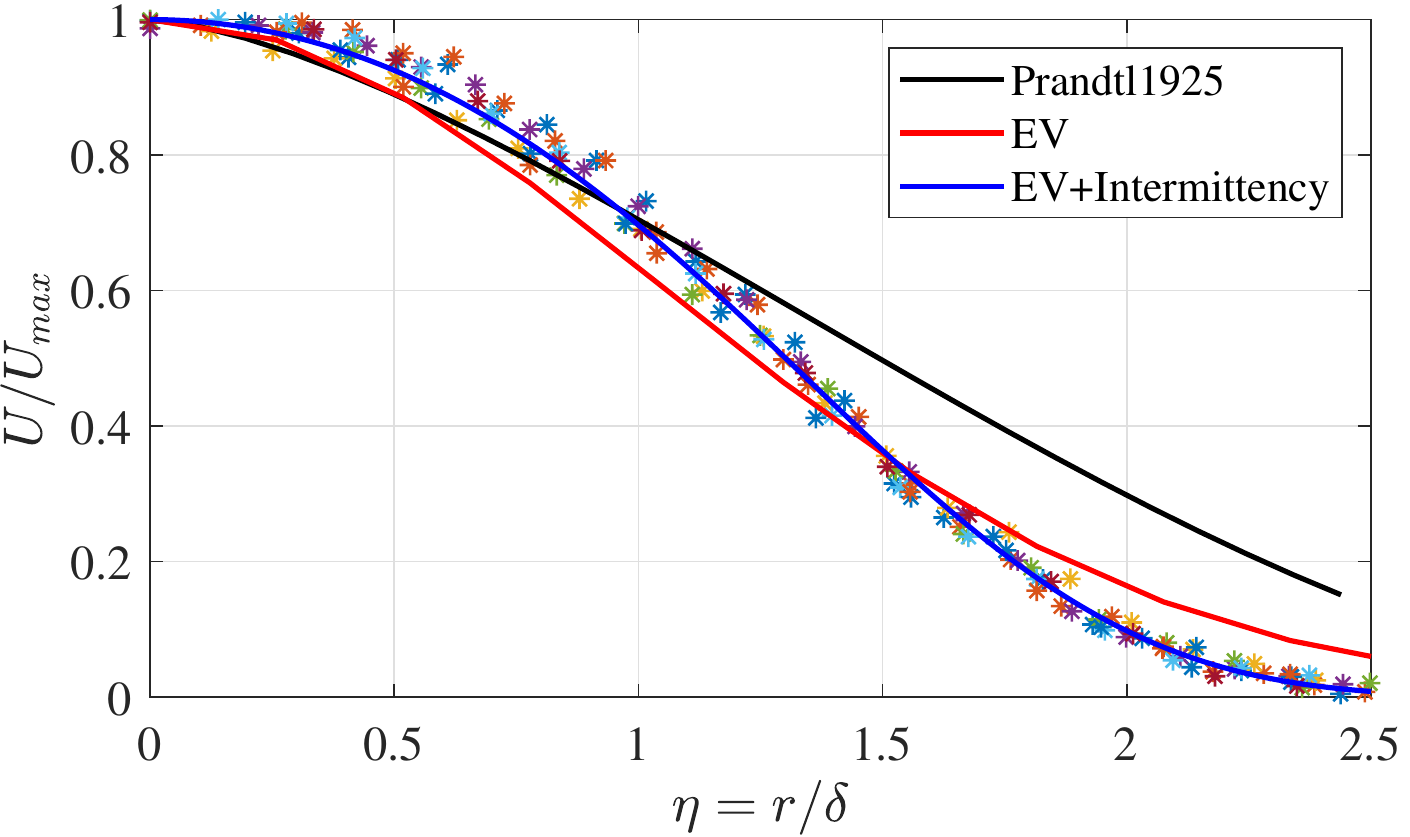}}
\subfloat[][]
{\includegraphics[width=0.5\columnwidth]{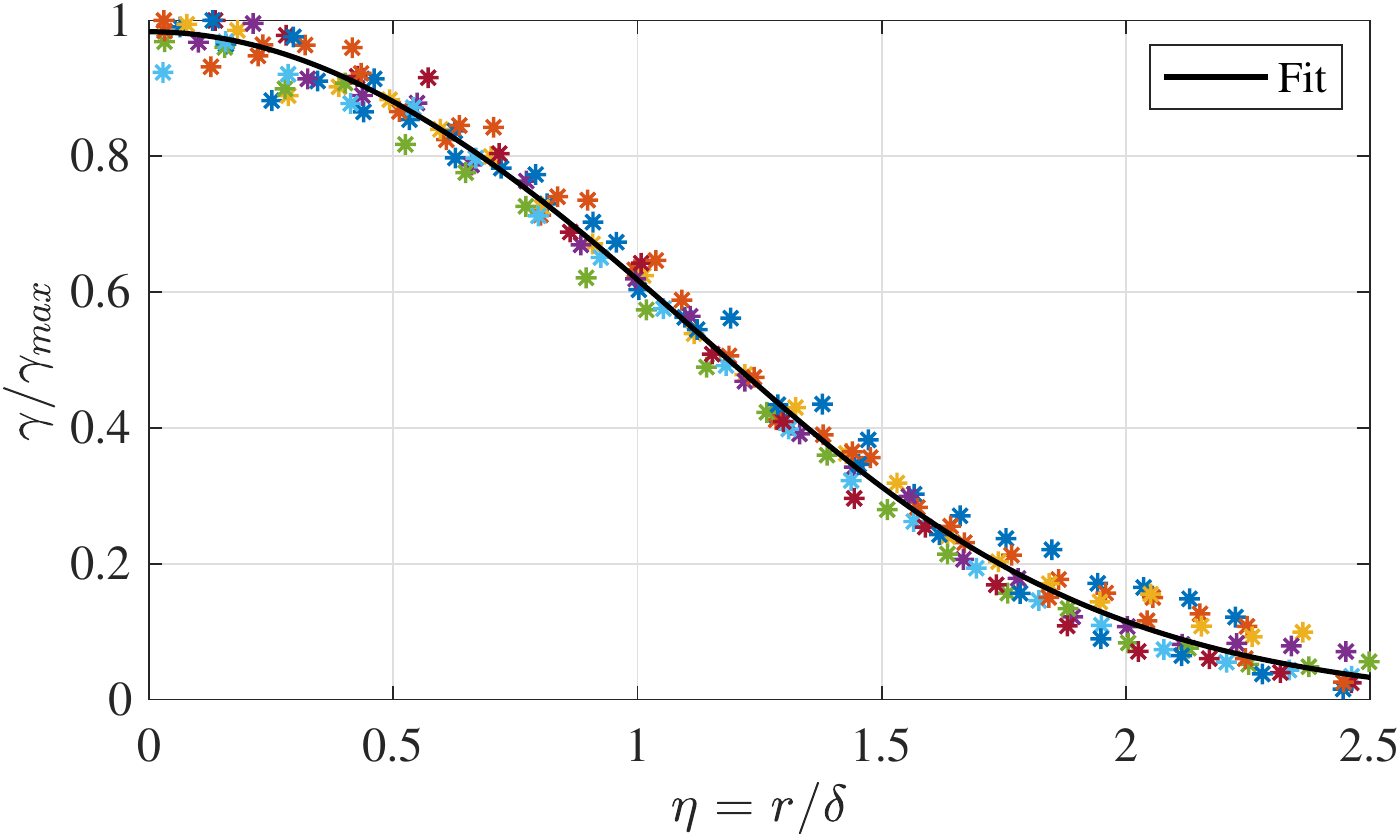}}
\caption{a) Mean flow velocity profiles for the axisymmetric wake in the range of streamwise distances $15 \leq x/L_{ref} \leq 50$ (symbols).  Black, red and blue lines are representative of equations \ref{eq:axiwakePML}, \ref{eq:axiwakeEV} and \ref{eq:TownProf}, respectively. b) Intermittency factor in the range of streamwise distances $15 \leq x/L_{ref} \leq 50$ rescaled with respect to its local maximum. The continuous line is representative of equation \ref{eq:gamma}. Data are plotted against the similarity coordinate $\eta=r/\delta$. The inlet Reynolds number is $Re_G=40000$.}
\label{fig:wakeprofiles}
\end{figure}
For an axisymmetric wake, when using Prandtl's mixing length (eq. \ref{eq:Mixinglength}) to model the Reynolds shear stress it is possible to show that the mean flow profile can be described as  \cite{goldstein} (see Appendix)
\begin{equation}
\frac{U}{U_{max}} = \sqrt{x} f(\eta) = \Bigl(1- \bigl(\frac{\eta}{\eta_0}\bigr)^{3/2}  \Bigr)^2,
\label{eq:axiwakePML}
\end{equation} 
{where $\eta=y/\sqrt{x}$ and $\eta=\eta_0$ ($y=y_0$) is the point where $f \rightarrow 0$ (already a non physical result) on the  boundary of the wake.}

When, instead, a constant value of the eddy viscosity is used (eq. \ref{eq:EddyViscosity}), the following form of the mean flow can be obtained \cite{johansson}:
\begin{equation}
\frac{U}{U_{max}} = e^{-\frac{k\eta^2}{2}},
\label{eq:axiwakeEV}
\end{equation} 
with $\eta = y/\delta$; $k$ depends on the turbulence dissipation
scaling (see Appendix). It is worth remarking that the profiles
obtained in equations (\ref{eq:axiwakePML})-(\ref{eq:axiwakeEV}) are
valid for both turbulence dissipation scalings (equilibrium and
non-equilibrium).

Figure \ref{fig:wakeprofiles}a shows the mean flow velocity profiles
rescaled with the maximum at each streamwise location plotted against
the similarity variable $r/\delta$. A comparison of the two proposed
models (eq. \ref{eq:axiwakePML}-\ref{eq:axiwakeEV}) suggests that
Prandtl's mixing length model significantly overestimates the velocity
profile for values of $r/\delta>1$. A constant value of the eddy
viscosity seems to follow more closely the physics of the problem. As
also showed by Townsend \cite{townsend1949}, a significant improvement
of the eddy viscosity model can be obtained by accounting for the
intermittency factor $\gamma$ calculated as the inverse of the
kurtosis of the time derivative of the streamwise fluctuating
velocity, \textit{i.e.},
\begin{equation}
\gamma = \frac{1}{\textup{Kurt}[d{u}^{\prime}/dt]}.
\label{eq:intermDef}
\end{equation}
This is due to the fact that the eddy viscosity cannot be non-zero beyond the turbulent/non-turbulent interface, where there are no vortical fluctuations at all. Following Townsend, we use a  functional form for the intermittency as
 \begin{equation}
\frac{\gamma}{\gamma_{max}} = \frac{1}{\bigl(1+(\eta/\alpha_1)^2+(\eta/\alpha_2)^4+(\eta/\alpha_3)^6\bigr)}
\label{eq:gamma}
\end{equation} 

\noindent with $\alpha_1$, $\alpha_2$ and $\alpha_3$ parameters to be
determined, and we redefine the eddy viscosity as $\nu_T^{I} = \gamma
\nu_T$. Equation (\ref{eq:gamma}) requires that the intermittency
  factor is self preserving; a condition satisfied in our experiments
  as can be observed from Figure \ref{fig:wakeprofiles}b where we plot
  the intermittency profiles obtained in the range of streamwise
  distances $10\leq x/L_{ref}\leq50$ normalised with respect to the
  local maximum at each streamwise location. The continuous line is
  representative of the fit of equation (\ref{eq:gamma}) and shows a
  remarkable agreement with the experimental data.

Hence, we modify the eddy viscosity accounting for the intermittency of the flow, \textit{i.e.} $\nu_T^I = \nu_T \gamma $, with $\nu_T$ a constant value and we find a solution to the self-similar equation of the form
 \begin{equation}
\frac{U}{U_{max}} =  e^{-{k\eta^2} {\bigl(\frac{1}{2}+\frac{1}{4}(\eta/\alpha_1)^2+\frac{1}{6}(\eta/\alpha_2)^4+\frac{1}{8}(\eta/\alpha_3)^6\bigr)}}.
\label{eq:TownProf}
\end{equation} 
We seek for the coefficients $\alpha_1$, $\alpha_2$, $\alpha_3$ and $k$ which optimise equations \ref{eq:gamma} and \ref{eq:TownProf} simultaneously. The continuous blue line in figure \ref{fig:wakeprofiles}a shows that the introduction of the intermittency factor significantly improves the results, particularly in proximity of the wake boundaries. Equation (\ref{eq:TownProf}) and variations of it have been extensively used to fit turbulent wakes \cite{Nedic2013} from bluff plates and wind turbines \cite{kermani2013} as \textit{ad hoc} modifications to a Gaussian profile.
\subsubsection{Planar Jet}
\begin{figure}
\centering
\subfloat[][]
{\includegraphics[width=0.5\columnwidth]{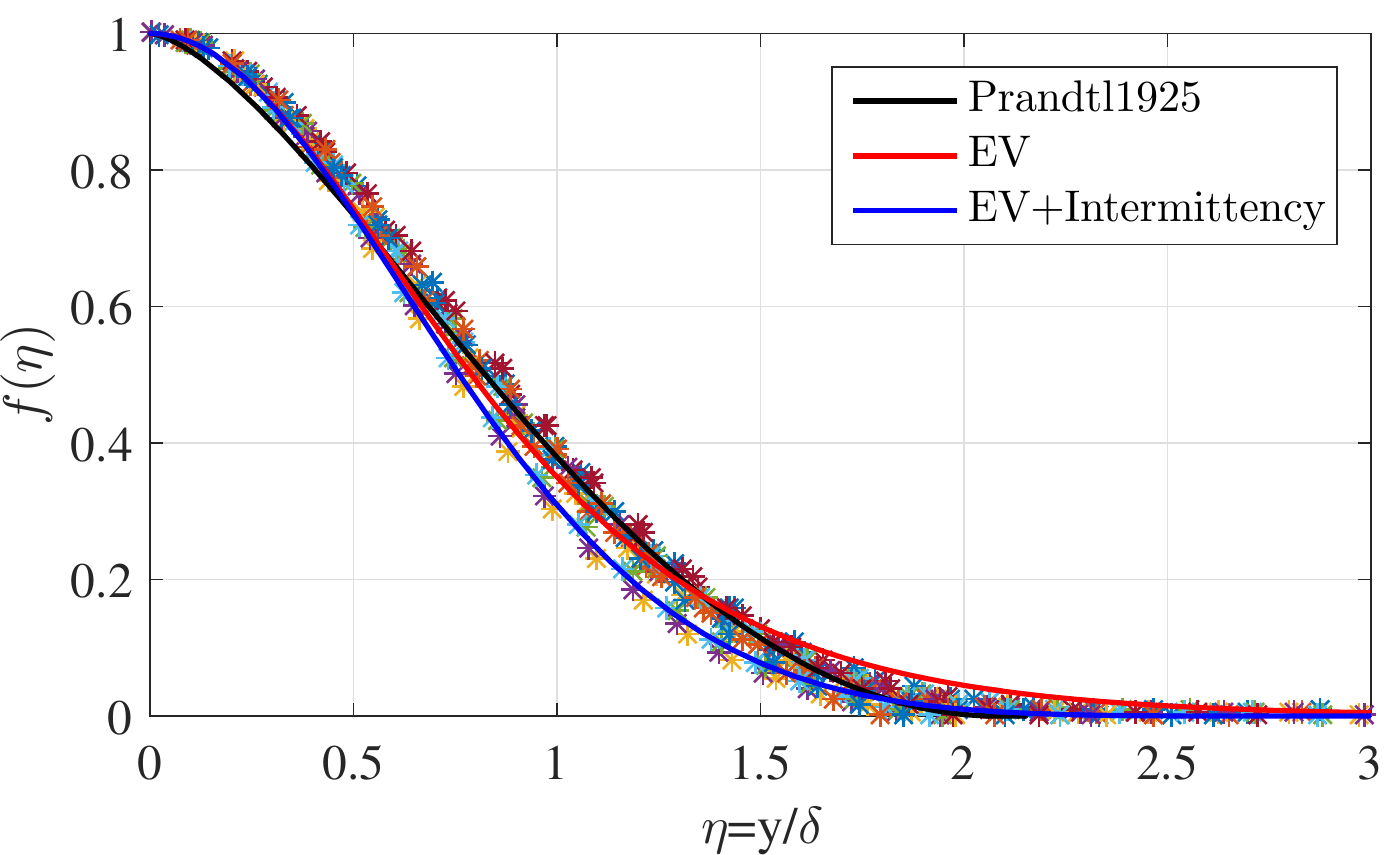}}
\subfloat[][]
{\includegraphics[width=0.5\columnwidth]{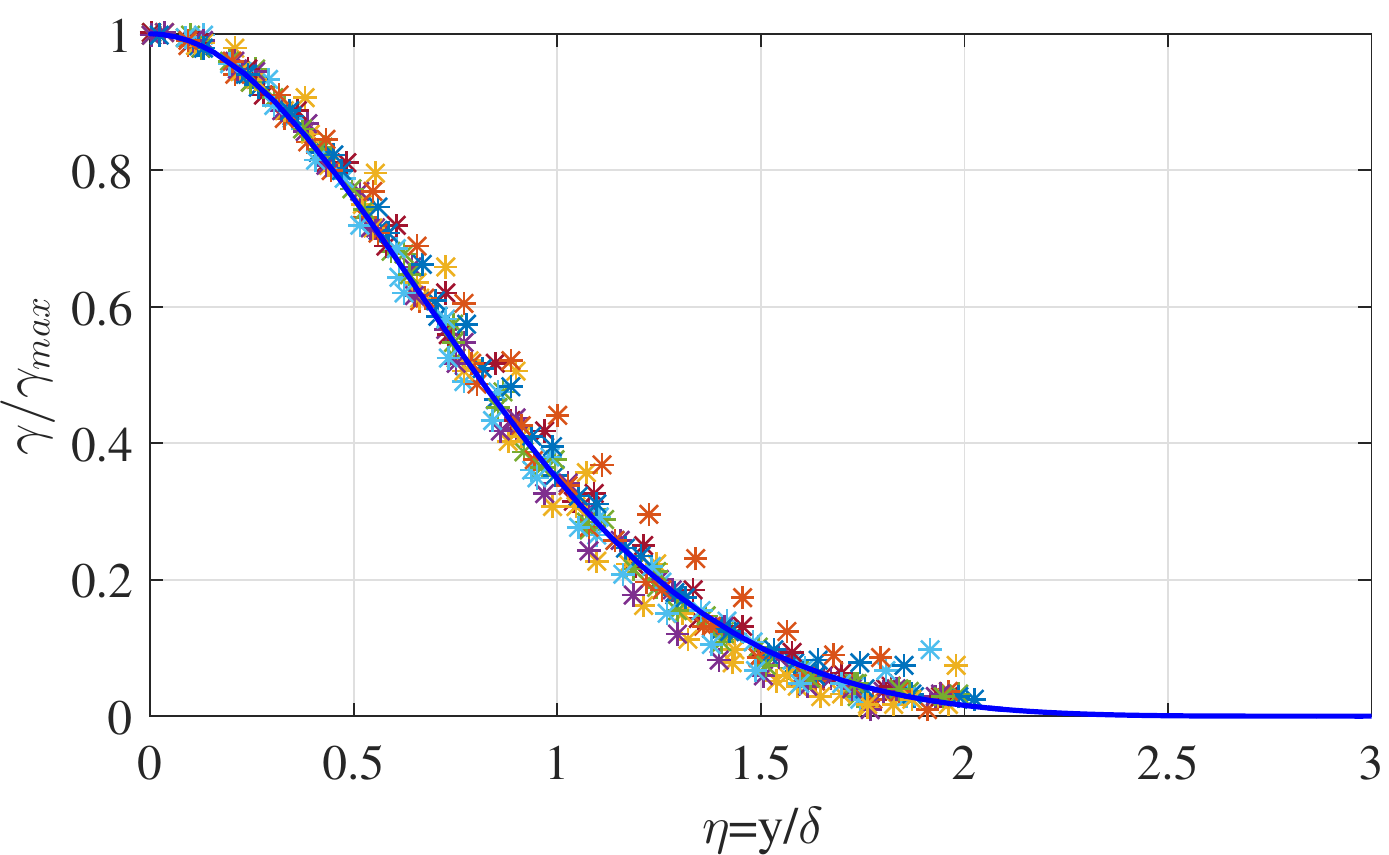}}
\caption{a) Mean flow velocity profiles for the turbulent planar jet in the range of streamwise distances $15\leq x/L_{ref}\leq 50$ (symbols).  Black and blue lines are representative of the solution obtained with Prandtl (1925), Eddy viscoity (EV) assumption or introducing the intermittency function, respectively. b) Intermittency function rescaled with respect to the local maximum at each streamwise position $x/L_{ref}$ in the range  $15 \leq x/L_{ref} \leq 50$. The black line is representative of the fit $\frac{\gamma}{\gamma_{max}}  = f(\eta)^m$. Data are plotted against the similarity coordinate $\eta=y/\delta$. The inlet Reynolds number is $Re_G=20000$.}
\label{fig:jetprof}
\end{figure}
We follow a similar procedure in the planar jet case. The application of Prandtl's mixing length model leads to the following form of the jet momentum equation
\begin{equation}
k{F^{\prime \prime}}^2 + F F^{\prime} = 0
\label{eq:MLJetProf}
\end{equation}
where $F^{\prime}(\eta) = f(\eta)$,
$f(\eta)=\overline{u}(x,y)/u_0(x)$, and $k$ a constant value which
depends on the turbulence dissipation scaling (this dependence is
reported in the Appendix). Nevertheless, as for the axisymmetric wake,
there is no substantial difference in the functional form of the mean
flow profile for the equilibrium and non-equilibrium cases. As
reported by Abramovich \cite{abramovich}, Tollmien \cite{tollmien} was
the first to find a numerical solution for equation
(\ref{eq:MLJetProf}). We also solve the equation numerically,
comparing the results with our experimental measurements and with the
results obtained with the eddy viscosity assumption with/without the
intermittency correction.

The adoption of the turbulent viscosity model, leads to the following momentum equation in similarity variables 
\begin{equation}
{F F^{\prime \prime}} + {F^{\prime}}^2 = 2k F^{\prime\prime\prime}
\label{eq:EVeq}
\end{equation}
(relations for $k$ are reported in the Appendix). This equation can be solved to obtain the velocity profile $f(\eta)$
\begin{equation}
f(\eta) = sech^2(\eta\sqrt{k}).
\label{eq:townJetProf}
\end{equation}
In Figure \ref{fig:jetprof}a we report the experimental data obtained from the planar jet experiment in the range of streamwise distances $15 \leq x/L_{ref}\leq 50$, along with the mean flow profiles predicted by the Prandtl mixing length model (black) and the eddy viscosity model (red). Despite very little differences throughout the whole range of lateral locations, it can be argued that the Prandtl model slightly underestimates the mean flow profile at small values of $\eta$. 

Even though  the eddy viscosity model shows good agreement with the experimental data, we decided to try and improve it by accounting for the intermittency of the flow, hence introducing $\nu_T^I = \nu_T \gamma$. As already discussed in the axisymmetric wake case, this requires that the intermittency function is self-similar. In Figure \ref{fig:jetprof}b we compare the data obtained in the range $15 \leq x/L_{ref}\leq 50$ rescaled with respect to the local maximum at each streamwise location ($\gamma_{max}$). It can be concluded that $\gamma$ is indeed self-similar. The momentum equation for the turbulent planar jet hence modifies as 
\begin{equation}
\Biggl(F'F  \Biggr)=2k\gamma F'',
\label{eq:momjetSS2text}
\end{equation}
where $F'(\eta)=f(\eta)$. The solution will depend on the choice of $\gamma$; we decide to introduce a function $\gamma(\eta)= f(\eta)^m$, with $m \neq 1$ ($m=1$ results in a sinusoidal function for $f(\eta)$) and we integrate numerically equation (\ref{eq:momjetSS2text}). The choice of a different intermittency function is mainly driven by the fact that equation (\ref{eq:gamma}) does not lead to a closed solution of equation (\ref{eq:momjetSS2text}), and we  considered a better choice by relating the intermittency function directly to the mean flow. This can also be instrumental in future investigations aimed at relating the intermittency function $\gamma$ to the turbulence cascade. As evidenced in Figure \ref{fig:jetprof}a the solution obtained with the introduction of the intermittency function (blue line) gives a slight improvement to the fit of the experimental data. 
\section{Conclusions}
Using experimental data from a turbulent planar jet and two
axisymmetric turbulent wakes, we find evidence for the proportionality
of the integral lengthscale $L$ and the characteristic flow width
$\delta$ in the range of streamwise distances $15 \leq x/L_{ref}\leq
50$. This is a fundamental hypothesis when deriving the self-similar
scaling laws in turbulent free shear flows, and it is now established
in the region where the turbulent dissipation scaling is of the now
known non-equilibrium type.

We then revisit the turbulence closure models such as Prandtl's mixing
length \cite{prandtl1925} and constant eddy viscosity
in the light of the non-equilibrium dissipation scalings. In this
framework, the cornerstone of Prandtl's (1925) model, \textit{i.e.}
$l_m \sim \delta $, is not valid. We show that $l_m \sim \delta
\sqrt{Re_G/Re_{0\delta}}$ instead. The scalings (\ref{eq:uo}) and
(\ref{eq:delta}) stemming from the non-equilibrium cascade agree with
the eddy viscosity model $\nu_T\sim U_{ref}L_{ref}$ rather than
$\nu_T\sim u_0\delta$, hence implying a constant value of the eddy
viscosity everywhere.

However, we demonstrate that these differences do not lead to
different mean flow profiles. A systematic comparison of the mean flow
profiles predicted by Prandtl (1925) and the eddy viscosity models
with the experimental data for the axisymmetric wake and the turbulent
planar jet reveals the inadequacy of Prandtl's mixing length
hypothesis to correctly predict the mean flow behaviour even where the
turbulence dissipation has a non-equilibrium cascade
scaling. Furthermore, following Townsend \cite{townsend1949} we show
that the prediction can be further improved by accounting for the
intermittency of the flow, particularly in the axisymmetric wake
case. In agreement with Townsend \cite{townsend1949}, we find that
rescaling the eddy viscosity $\nu_T$ with the intermittency function
$\gamma$ provides a better representation of the mean flow behaviour
as it accounts for the eddy viscosity drop across the turbulent flow's
intermittent boundaries.
\section*{Acknowledgements}
GC and JCV were supported by ERC Advanced Grant 320560 awarded to JCV.
 \section*{Appendix A: governing equations}\label{sec:app}
In this Appendix we report about the equations that lead to the
definition of the lateral profiles for the axisymmetric wake and the
planar jet flows, following \cite{dairay2015} and \cite{cvjet}
respectively. We start by describing the axisymmetric wake, then
particularise the radial profiles according to the different closure
models for the Reynolds stresses.
\subsubsection*{Axisymmetric wake}
\noindent In the thin shear layer approximation, the momentum balance reads
\begin{equation}
U_{ref} \frac{\partial}{\partial x}(U_{ref}-U) = -\frac{1}{r}\frac{\partial}{\partial r}(r \langle u'v'\rangle).
\label{eq:momwake}
\end{equation}
Assuming that $U_s= U_{ref}-U= u_0 f(\eta)$, and substituting in equation (\ref{eq:momwake}) 
\begin{equation}
U_{ref} \frac{\partial}{\partial x}(f(\eta)) - U_{ref}U_s f'(\eta)\eta\frac{1}{\delta}\frac{\partial}{\partial x} \delta  = -\frac{1}{r}\frac{\partial}{\partial r}(r \langle u'v'\rangle).
\end{equation} 
Momentum flux constancy $U_s\delta^2 = U_{ref}\theta^2$ can be differentiated to get 
\begin{equation}
\frac{\partial}{\partial x} U_s = -2 \frac{U_s}{\delta}\frac{\partial}{\partial x}\delta, 
\end{equation} 
so we rewrite the momentum equation as
\begin{equation}
\frac{2U_{ref}U_s}{\delta}\frac{\partial}{\partial x}(\delta) f(\eta) + U_{ref}U_s f'(\eta)\eta\frac{1}{\delta}\frac{\partial}{\partial x} \delta  = \frac{1}{r}\frac{\partial}{\partial r}(r \langle u'v'\rangle).
\label{eq:momFin}
\end{equation} 
The solution of equation (\ref{eq:momFin}) depends on the Reynolds stress modelling. As reported by Goldstein \cite{goldstein}, the classical streamwise dependent eddy viscosity based on Prandtl's mixing length (Prandtl (1925)),
\begin{equation}
\nu_T = l_m^2\frac{\partial}{\partial r}(U_{ref}-U),
\end{equation}
leads to the following equation,

\begin{equation}
f'=\sqrt{\left( \frac{\eta f}{k} \right)},
\label{eq:axiwakePMLApppartial}
\end{equation} 

with $k=\frac{l_m^2U_s}{\delta^2\delta'U_{ref}}$. This equation has the solution,

\begin{equation}
\frac{U}{U_{ref}} =  f(\eta) = \Bigl(1- \bigl(\frac{\eta}{\eta_0}\bigr)^{3/2}  \Bigr)^2,
\label{eq:axiwakePMLApp}
\end{equation} 
{where $\eta_0=(3k)^{1/3}$ is the point at the  boundary of the wake (and therefore where $f \to 0$).}  Now, depending on the properties of the turbulent cascade, two different closures can be obtained:

\textbf{i)} Richardson Kolmogorov cascade:

In this case, we have $l_m=C \delta$, with $C$ a
constant. Furthermore, the streamwise scalings are $u_0=
AU_{ref}(\frac{x-x_0}{\theta})^{-2/3}$ and $\delta=
B\theta(\frac{x-x_0}{\theta})^{1/3}$. Adding the integral form of
momentum conservation ($u_0 \delta^2=U_{ref} \theta^2$), we get that
$k=3\frac{C^2}{B^3}$ and $\eta_0=9\frac{C^{2/3}}{B}$.

\textbf{ii)} Non-equilibrium cascade:

In this case, we have $l_m=C \delta \sqrt{Re_G/Re_{0\delta}}=C
\sqrt{U_{ref} L_{ref} \frac{\delta}{U_s}}$, and again $C$ is a
constant. In this case, the streamwise scalings are $u_0=
AU_{ref}\left(\frac{x-x_0}{L_{ref}}\right)^{-1}(\theta/L_{ref})^2$ and
$\delta= B\sqrt{L_{ref} (x-x_0)}$. Therefore, the constant becomes
$k=2\left(\frac{C}{B}\right)^2$ and
$\eta_0=6\left(\frac{C}{B}\right)^{2/3}$.

\noindent Conversely, the adoption of a turbulent eddy viscosity
model, $\nu_T=constant$ delivers a substantially different lateral
velocity profile

\begin{equation}
\frac{U}{U_{ref}} = e^{-\frac{k\eta^2}{2}},
\label{eq:axiwakeNEQApp}
\end{equation} 
with $\eta = y/\delta$ and $k =\frac{U_{ref} \delta  \frac{d\delta }{dx}}{\nu_T}$. Once more, two different closures can be obtained:

\textbf{i)} Richardson Kolmogorov cascade:

We have $\nu_T=C u_0 \delta$, with $C$ constant, and we find that $k=\frac{1}{3}\frac{B^3}{C}$. 

\textbf{ii)} Non-equilibrium cascade:

In this case, we have $\nu_T=C U_{ref} L_{ref}$. The constant becomes $k=\frac{1}{2}\frac{B^2}{C}$. 

On the other hand, Townsend \cite{townsend1949}, studying the planar wake, suggested that the quality of the fit could be further improved by accounting for the intermittency of the flow $\gamma$. He proposed the use of $\gamma$ in a modified eddy viscosity $\nu_T^{I} = \gamma \nu_T$ where

 \begin{equation}
\frac{\gamma}{\gamma_{max}} = \frac{1}{\bigl(1+(\eta/\alpha_1)^2+(\eta/\alpha_2)^4+(\eta/\alpha_3)^6\bigr)},
\label{eq:gammaApp}
\end{equation} 
leading to the following correction of equation (\ref{eq:axiwakeNEQApp}) 
 \begin{equation}
\frac{U}{U_{max}} =  e^{-{k\eta^2} {\bigl(\frac{1}{2}+\frac{1}{4}(\eta/\alpha_1)^2+\frac{1}{6}(\eta/\alpha_2)^4+\frac{1}{8}(\eta/\alpha_3)^6\bigr)}},
\label{eq:TownProfApp}
\end{equation} 

\noindent where $k$ remains unchanged from the previous case, not corrected by the intermittency.

\subsubsection*{Planar Jet}
\noindent In the thin shear layer approximation, the streamwise momentum equation for the planar jet flow  is
\begin{equation}
U \frac{\partial}{\partial x}(U)+V \frac{\partial}{\partial y}(U) = -\frac{\partial}{\partial y}( \langle u'v'\rangle),
\label{eq:momjet}
\end{equation}
\noindent and similarly to the wake flow, we can assume that the mean flow is self similar $U=u_0 f(\eta)$. This condition, along with continuity, 
\begin{equation}
\frac{\partial}{\partial x}(U)+\frac{\partial}{\partial y}(V) = 0,
\label{eq:contjet}
\end{equation}
implies that  the lateral velocity is self-similar as well. Casting equations (\ref{eq:momjet}) and (\ref{eq:contjet}) together and using the self similarity of the mean flow we obtain
\begin{equation}
\frac{\partial}{\partial x}(\delta) \Biggl(f^2+\frac{f'}{\eta} \int_0^{\eta} f(\overline{\eta}) d\overline{\eta}   \Biggr)=-\frac{\partial}{\partial y}( \langle u'v'\rangle).
\label{eq:momjetSS}
\end{equation}
Introducing $F=f'$ and rearranging the equation, we get
\begin{equation}
\frac{\partial \delta}{\partial x}\Biggl(F'F  \Biggr)=-\frac{\partial}{\partial y} u^{\prime}v^{\prime}.
\label{eq:momjetF}
\end{equation}
Equation (\ref{eq:momjetF}) is then particularized depending on the closure for the Reynolds shear stresses. Prandtl's mixing length model leads to the following equation,
\begin{equation}
k{F^{\prime \prime}}^2 + F F^{\prime} = 0,
\label{eq:momjetPMLApp}
\end{equation}
with $k=2\frac{l_m^2}{\delta^2\delta'}$. This equation has no known
analytical solution, hence we solve it numerically. We can again
relate the constant $k$ to model constants depending on the type of
turbulence cascade:

\textbf{i)} Richardson Kolmogorov cascade:

The mixing length is $l_m=C \delta$, with $C$ constant. The streamwise scalings are $u_0= AU_j(\frac{x-x_0}{h})^{-1/2}$ and $\delta= Bh(\frac{x-x_0}{h})$, and therefore we get that $k=2C^2/B$. 

\textbf{ii)} Non-equilibrium cascade:

In this case, we have $l_m=C \delta \sqrt{Re_G/Re_{0\delta}}$, and
again $C$ is a constant. In this case, the streamwise scalings are
$u_0= AU_{ref}\left(\frac{x-x_0}{h}\right)^{-1/3}$ and $\delta=
BL_{ref}\left(\frac{x-x_0}{h}\right)^{2/3}$. Adding the integral form
of momentum conservation ($u_0 ^2\delta=U_j^2 h$), the constant
becomes $k=3\sqrt{\frac{C}{B^3}}$.

Assuming now a constant turbulent eddy viscosity model, the momentum
equation particularizes as follows
\begin{equation}
\Biggl(F'^2+F''F  \Biggr)=k F''',
\label{eq:momjetSS1}
\end{equation}
with $k =\frac{\delta  u_0 \frac{d\delta}{dx}}{2 \nu_T}$. Finally,

\begin{equation}
\Biggl(F'F  \Biggr)'=k\gamma F''.
\label{eq:momjetSS2}
\end{equation}

\noindent This equation can be solved to obtain the velocity profile
$f(\eta)=F'(\eta)$,
\begin{equation}
f(\eta) = sech^2(\eta \sqrt{k} ).
\label{eq:townJetProfApp}
\end{equation}

Again, depending on the properties of the turbulent cascade, both
$\nu_T$ and $k$ will adopt different values:

\textbf{i)} Richardson Kolmogorov cascade:

The eddy viscosity takes the form $\nu_T=C u_0 \delta$ with $C$ a
constant. Therefore, we get $k=2CA^2$.

\textbf{ii)} Non-equilibrium cascade:

We have $\nu_T=C U_{ref} L_{ref}$, and the constant becomes $k=3CA^3$. 

\noindent Similarly to the axisymmetric wake case, we also study the
case of modified eddy viscosity $\nu_T^{I} = \gamma \nu_T$. Equation
(\ref{eq:momjetSS1}) then becomes
\begin{equation}
\Biggl(F'F  \Biggr)=k\gamma F'',
\label{eq:momjetSS2}
\end{equation}
whose solution depends on the choice of $\gamma$. We propose a
function $\gamma=(f(\eta))^m$ (with $m \neq 1$), relating the
intermittency to the mean flow profile. As there is no known closed
solution, we numerically solve equation (\ref{eq:momjetSS2}).

\bibliography{jfm-instructions}

\end{document}